\documentclass[a4paper,journal]{IEEEtran}
\usepackage{amsfonts}
\usepackage[fleqn]{amsmath}
\usepackage{algorithmic}
\usepackage{array}
\usepackage[caption=false,font=normalsize,labelfont=sf,textfont=sf]{subfig}
\usepackage{textcomp}
\usepackage{stfloats}
\usepackage{url}
\usepackage{verbatim}
\usepackage{graphicx}
\def\BibTeX{{\rm B\kern-.05em{\sc i\kern-.025em b}\kern-.08em
		T\kern-.1667em\lower.7ex\hbox{E}\kern-.125emX}}
\usepackage{balance}

\usepackage{cite}

\usepackage{multirow}
\usepackage{tabularx,booktabs}

\usepackage{ragged2e}

\usepackage[flushleft]{threeparttable}

\usepackage{siunitx}
\usepackage{enumerate}

\usepackage{diagbox}

\usepackage[euler]{textgreek}

\usepackage{mathtools}
\usepackage{url}

\usepackage{float}
\usepackage{array}

\usepackage{bigdelim}

\usepackage[flushleft]{threeparttable}

\usepackage{color,soul}

\usepackage{xcolor}

\begin{document}
\title{Inductance Estimation for High-Power Multilayer Rectangle Planar Windings}
\author{\IEEEauthorblockN{Theofilos Papadopoulos,
		Antonios Antonopoulos}
		\IEEEauthorblockA{\\School of ECE,
		NTUA\\
		Zografou, Greece\\
		Email: teopap@mail.ntua.gr}}


\maketitle

\begin{abstract}
This paper proposes a simple and accurate monomial-like equation for estimating the inductance of Multilayer Rectangle-shaped Planar Windings (MLRPWs) for high-frequency, high-power applications. The equation consists of the power product of the geometrical dimensions, raised at individual power coefficients. The coefficients are generated via Multiple Linear Regression (MLR), based on a large set of approximately 6,000 simulated windings, with an 80/20 training/evaluation sample ratio. The resulting mean error value is \textmu=0\%, with a standard deviation below 1.8\%. The accuracy of the inductance estimation is confirmed on several experimental samples, with dimensions both within and outside the initial training dataset.


\end{abstract}


\section{Introduction}

Current trends in power applications are raising the demand for high-frequency (HF) converters in order to increase power density, without compromising high efficiency figures \cite{HF_PES}. Increasing the switching frequency of the power converters can lead to solutions of greatly reduced size (for the same power-handling capability);
however, special care needs to be taken on the design of the magnetic components. Based on their manufacturing method two main groups of magnetic components (inductors and transformers) can be identified: conventional architectures based on conductors wrapped around an axis (or core) and planar architectures extending only in xy-plane (instead of z-axis).
The latter group, originally associated with radio-frequency (RF) designs, is emerging for high-power high-frequency applications, offering various advantages, compared to conventional designs \cite{yao23, cao23}.

Making use of their geometrical deployment, planar windings (PWs) can be utilized in applications where the height of the components is constrained, such as rack-mounted power supplies in data centers \cite{DataCenter}, electric vehicle charging converters \cite{ev1,IPT2}, wireless power transfer \cite{IPT}, solid-state DC transformers \cite{rev1_dong_dcx}, or even in aircraft propulsion converters \cite{rev1_dong_propulsion}. Due to their relatively large surface area, PWs usually offer good thermal characteristics as well \cite{thermal1}. Additionally, since the copper height in PCBs is typically 35 \textmu m (or less), there is no need for skin-effect compensation techniques for frequencies up to 3 MHz \cite{saturn}. In case skin-effect compensation is deemed necessary, a way to design Litz-wire equivalent traces is suggested in \cite{Litz21}. Moreover, a great advantage of PWs is that they can be effectively produced using PCB manufacturing processes, with very low manufacturing costs. At the same time, this process provides well-determined inductance with extremely low deviation from the desired value. This property is especially useful for soft-switching converters \cite{cllc1, cllc2, llc}, where the exact knowledge of the inductance value and the capability to reproduce elements with the exact same properties are crucial for proper control and increased efficiency. 

Estimating, however, the inductance of a PW is not always a straightforward task, since the implemented shape (e.g. square, oval, polygons, multilayer, etc.) affects the inductance value in different ways \cite{param_est, leak_param, review}. The most accurate estimations can be achieved through analytical solutions \cite{aebischer2020}, which result in relatively complex expressions, difficult to adapt for different dimensions. Furthermore, in order to simplify the derivation of such expressions, the assumption of regular polygons is used, limiting their applicability to other shapes, and as asymmetries (deviation from regular shapes) becomes dominant to better utilize available space, these solutions become ineffective.

In the effort for enhanced space utilization, PW designs can be implemented in multilayer architectures, via proper folding of similar single-layer windings, so that the magnetic flux of each layer is generated in the same direction. These designs can increase the inductance exponentially, while achieving a high coupling factor and maintaining the z-profile low. Additionally, further benefits can be achieved by lifting the restriction for ``regular-polygon'' shape symmetry. Rectangle-shaped PWs (RPWs) provide the designer with one more degree of freedom, which can be crucial in applications with limited space availability.  



In the absence of closed-form expressions, the standard method for calculating the inductance value of PWs is to use iterative methods and time-consuming finite-element model (FEM) simulations. This process complicates the design of power converters, as the values of passive elements have an impact on the determination of many other design parameters, such as operating frequency of resonant converters, the ripple current of DC/DC converters, the power transfer capabilities of active-bridge converters, etc. In this respect, it is favorable for designers of high-power, high-frequency applications to be able to determine the exact inductance of a PW quickly, without the need to revert to iterative methods or FEM simulations for each change they apply during the design process. This can be achieved through closed-form expressions relating the geometrical dimensions of PWs to their inductance value. Historically, such expressions were developed for single-layer PWs of regular-shapes, and aiming for RF applications \cite{wheeler, ssmohan}, which present significantly different characteristics compared to the demands of high-power applications. 

Previous work has shown that by modifying the equations for single-layer windings it is possible to expand the inductance estimation from square-shaped to rectangular-shaped windings \cite{Papadopoulos23}. 
Also, the approximation of substituting the number of turns with the product of the turns and the number of layers has been evaluated in efforts to provide a simple estimation method for multilayer square-shaped PWs. This technique provides good results for windings up to two layers, but shows low accuracy when the number of layers is larger \cite{Papadopoulos22}. The main reason for the loss of accuracy is the increased leakage flux and the reduced mutual inductance between layers, when their distance is increasing. It is obvious that the distance between layers has to be considered when estimating the overall inductance. However, introducing a new parameter in empirical expressions  derived for significantly different winding architectures is not a straight-forward task. On the other hand, the Monomial equation is more suitable for expanding into multiple layers, as, due to its form, it can accommodate the introduction of new geometrical parameters with the proper redetermination of the exponential coefficients.
 
This study presents a novel new form of the Monomial-like equation, initially presented in \cite{ssmohan}, towards both multilayer and rectangular shape high-power PWs (MLRPWs). The result is a simple monomial-like expression that can estimate the inductance based only on the geometrical dimensions of the MLRPW. This enables the exact determination of the inductance and paves the way for fast spatial optimization of the winding. Additional parameters are considered compared to \cite{ssmohan}, in order to adapt to different dimensions.

In order to achieve enhanced estimation accuracy, the power coefficients are extracted via multiple linear regression (MLR) from a dataset containing more than 5,800 samples. The validity of the equation is confirmed by laboratory measurements for frequencies up to 50 kHz, even for windings with dimensions outside the training dataset.



This paper is organized as follows: In Section \ref{Sec:ParamDisc}, the original Monomial equation for single-layer square-shaped windings, as presented in \cite{ssmohan}, is discussed. The role and the weight of each parameter is analyzed. In Section \ref{Sec:EqForm}, the form of the suggested extension is presented, along with the methodology to estimate the correct coefficients. The experimental setup and the results that verify the accuracy of the estimation formula are presented in Section \ref{Sec:ExpVerif}, and in Section \ref{Sec:Conclusion} the final conclusions are drawn.

\section{Original Equation and Parameter Discussion} \label{Sec:ParamDisc}

The original Monomial equation, as presented in \cite{ssmohan}, has the form

{\setlength{\mathindent}{0cm}
\begin{equation}
	\label{Eq:mn_original}
	L_{\text{MN}} = 1.62 \cdot 10^{-3} D^{-1.21} \left( \frac{D+d}{2} \right) ^{2.4} w^{-0.147} s^{-0.03} N^{1.78},
\end{equation}

\noindent
where $D$ the outer-side length, $d$ the inner-side length, $w$ the width of the copper trace, $s$ the spacing between two consecutive traces and $N$ is the number of turns, as presented in Fig. \ref{Fig:Planar_1}. All geometrical parameters are measured in \textmu m and the resulting inductance is in nH. Eq. \eqref{Eq:mn_original} can be rewritten as

\begin{equation}
	\label{Eq:mn}
	L_{\text{MN}} = 1.5428 \mu_0 D^{-1.21} \left( \frac{D+d}{2} \right) ^{2.4}  w^{-0.147} s^{-0.03} N^{1.78},
\end{equation}

\noindent
where all variables are measured in SI units. The constant term $1.62 \cdot 10^{-3}$ has been changed to $1.5428 \mu_0$, while the rest of terms remain the same. Eq. \eqref{Eq:mn} can be rewritten in a more compact form as 

\begin{equation}
	\label{Eq:mn_2}
	L_{\text{MN}} = 1.5428 \mu_0 D^{-1.21} \overline{D}^{2.4} w^{-0.147} s^{-0.03} N^{1.78},
\end{equation}

\noindent
where $\overline{D} = \frac{D+d}{2}$.

The geometrical parameters are not fully independent, as $D$, $N$, $w$ and $s$ dictate the value of $d$ as in

{\setlength{\mathindent}{2.5cm}
\begin{equation}
	\label{Eq:d}
	d = D - 2 N (w+s) + 2s.
\end{equation}

\noindent
Obviously, $d$ cannot be negative, and the minimum desired value defines the external side length as well. The track width $w$ and spacing $s$ can be determined by the standards IPC-2152 for the current and IPC-2221 for the voltage  handling capability of the winding. 

According to the power coefficients in (\ref{Eq:mn_2}), the variable that affects the inductance the most is the mean value of the perimeter of the winding $\overline{D}=\frac{D+d}{2}$, which is raised to 2.4. The importance of this parameter was already identified in the first estimation attempts \cite{rosa}. It derives from the Current Sheet Approximation (CSA), where copper traces on the same side (being close to each other and carrying current in the same direction) create a strong positive mutual inductance. Accordingly, traces on opposite sides carry current in opposing directions, generating a negative mutual inductance. The strength of the negative mutual inductance decreases as the distance between the two sides increases, as for CSA the two sides of the winding can be regarded as two conductors with opposite currents. This reveals the important role of $d$ when considering the total inductance of a winding. 

The outer-side length $D$ is present in the numerator (as a sum) raised to 2.4 and in the denominator raised to 1.21, hence the total inductance increases with $D$. The number of turns $N$ also has a strong impact on inductance, as it is raised to 1.78. Therefore, the variables $D$, $d$ and $N$ play the most significant role in the determination of the total inductance. 

The inductance decreases with the width of the copper $w$ (raised to -0.147), since (with $D$ and $N$ constants) larger $w$ results to smaller $d$. The impact of the spacing between traces $s$ is almost negligible, at least for the dimensions this paper considers. In practice, the spacing between the traces is not an important design issue, since a few \textmu m of clearance, especially when coating is present, can isolate more than 100 V (IPC-2221B, B4 external conductors with polymer coating \cite{saturn}). The voltage distribution along each turn is briefly discussed in \cite{Papadopoulos21}.

\begin{figure}[!t] 
	\label{Fig:Planar} 
	\centering
	\subfloat[\label{Fig:Planar_1}]{%
		\includegraphics[width=2in]{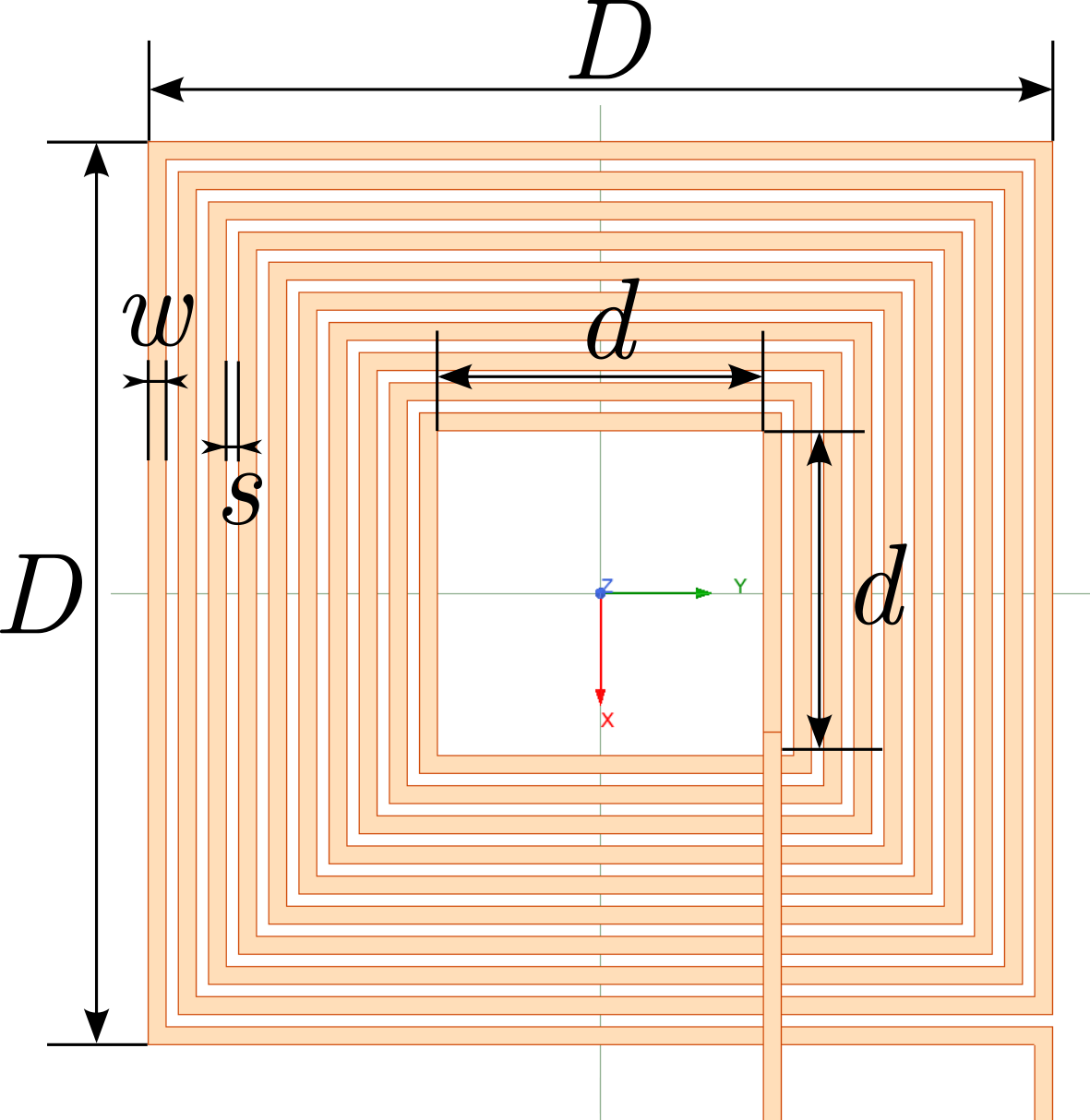}}
	\\
	\subfloat[\label{Fig:Planar_2}]{%
		\includegraphics[width=2.5in]{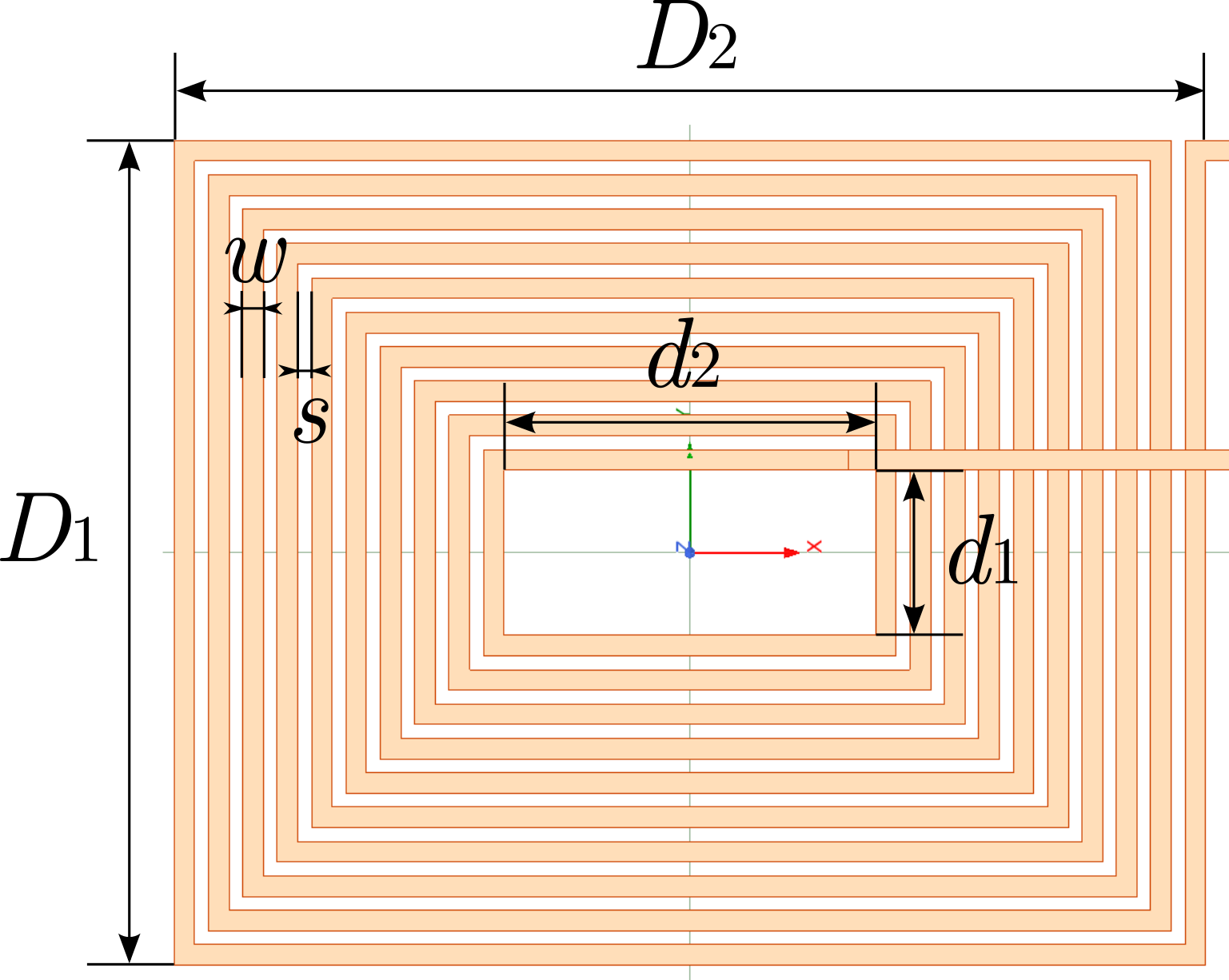}}
	\caption{Geometrical parameters of planar windings of (a) square shape and (b) rectangle shape.}
\end{figure}

In conclusion, as it is intuitively expected, larger external dimensions (larger $D$) increase the inductance. The inductance also increases by adding more turns (larger $N$) which, for a constant $D$, can be done either by decreasing $w$ or $d$. In the first case the inductance will exponentially increase as long as $d$ is kept approximately the same. However, reducing $w$ may have adverse effects on the winding temperature, which will increase as well, since the same amount of current will run through a narrower trace. In case $d$ is decreased, it is uncertain if the inductance will increase, since $\overline{D}$ gets decreased (traces carrying currents in opposite directions get closer). 

The Monomial equation in its current form, as given in (\ref{Eq:mn_2}), does not allow estimations for windings with different side lengths, or for multiple layers (properly) connected in series to constitute a single winding. However, keeping the same assumptions for the aforementioned parameters, can provide a way to extend the estimation to other windings forms as well. 

The Monomial equation is preferred for the specific task due to its easy expandability compared to other equations (e.g. Wheeler's or Rosa's). New variables (geometrical dimensions) can be added in the product of the equation along with a corresponding exponential coefficient. The value of the coefficient can be determined by algorithms (e.g., multiple linear regression) that minimizes the error between a given dataset and the equation.




\section{Simulation Set, Equation Form and MLR} \label{Sec:EqForm}

The Monomial equation is easily expandable to include new variables. Furthermore, it has been shown in \cite{Papadopoulos21, Papadopoulos22} that a recalculation of its coefficients may be necessary to achieve good accuracy for dimensions relevant to high-power applications. Several modifications are required to adapt the Monomial equation for estimation of rectangular and also multilayer PWs. Firstly, $D$ and $d$ need to be separated into  $D_1$, $D_2$ and $d_1$, $d_2$, respectively, as presented in Fig. \ref{Fig:Planar_2}. Moreover, two new parameters need to be introduced: the number of layers $N_L$ and the vertical distance between two consecutive layers $O$. 



High-power MLRPWs require traces with relatively large width $w$, leading to also relatively large outer-side lengths $D_1$ and $D_2$. For this study, $w$ is selected from 3 mm to 5 mm, which corresponds to current roughly from 5 to 10 A (for a maximum increase in temperature of $20~^\text{o}$C \cite{saturn}). Spacing $s$ varies from 0.1 to 0.5 mm, so that the voltage-withstand capability between two consecutive turns, is up to 300 V (for B4 external conductors with coating). The number of turns per layer $N = N_T$ varies from 6 to 10 turns, as they are used quite often in real windings, and the number of layers $N_L$ varies from 1 to 4, for the same reason. The inner-side lengths $d_1$ and $d_2$ are limited to be greater or equal than 17 mm, in order keep the two sides apart and maintain relatively high inductance, as described in Section \ref{Sec:ParamDisc}. Allowing smaller values on $d_1$ and $d_2$ would lead to a reduction of the total inductance. This limitation allows for a wide-enough aperture that can also accommodate for a central core leg (EE or EI). The aforementioned values for each variable are presented aggregated in Table \ref{Table:Datasets}.

\begin{table}[!tb]
	\centering
	\caption{Simulated Datasets for MLR Training}
	\label{Table:Datasets}
	\begin{tabular}{lcc}
		\toprule
		& Dataset A                                                                                           & Dataset B               \\ \midrule
		$D_1$ & \begin{tabular}[c]{@{}l@{}}70:10:110 mm\\ $D_1 \le D_2$\end{tabular} & \begin{tabular}[c]{@{}l@{}}120:10:160 mm\\ $D_1 \le D_2$\end{tabular} \\ 
		$D_2$ & \begin{tabular}[c]{@{}c@{}}70:10:110 mm\end{tabular}                          & \begin{tabular}[c]{@{}c@{}}120:10:160 mm\end{tabular}                          \\ 
		$d_1$ & \multicolumn{2}{c}{\begin{tabular}[c]{@{}c@{}}from (\ref{Eq:d}), $d_1$ \textgreater 17 mm\end{tabular}}                      \\ 
		$d_2$ & \multicolumn{2}{c}{\begin{tabular}[c]{@{}c@{}}from (\ref{Eq:d}), $d_2$ \textgreater 17 mm\end{tabular}}                        \\ 
		$w$  & \multicolumn{2}{c}{3, 4, 5 mm}                          \\ 
		$s$  & \multicolumn{2}{c}{0.1, 0.3, 0.5 mm}                                                                                    \\ 
		$N_T$ & \multicolumn{2}{c}{6, 8, 10 turns}                                                                                             \\ 
		$N_L$ & \multicolumn{2}{c}{1, 2, 3, 4 layers}                                                                                          \\ 
		$O$  & \multicolumn{2}{c}{\begin{tabular}[c]{@{}c@{}}0.5, 1.0, 1.5 mm\\ not defined for $N_L=1$\end{tabular}}                      \\ \bottomrule                            
	\end{tabular}
\end{table}


As the orientation of the winding is irrelevant, meaning that the inductance for a winding with dimensions \{$D_1^*,D_2^*$\} is the same with \{$D_2^*,D_1^*$\}, the restriction $D_1 \le D_2$ is applied and reduces the necessary simulations by a factor of 2. 
Considering that a large number of FEM simulations is a time-consuming task (in the order of several months in the case of this research), another reduction by a factor of two is favorable. In  order to keep the dimensions range, but avoid unnecessarily many combinations, the dimension subset C $\{D_1,D_2\}=\{70:110,120:160\}$ has been omitted. This further reduces the total simulation time that is required to generate the training dataset, but without reducing the overall accuracy of the equation.

\begin{figure}[!t] 
	\centering
	\includegraphics[width=2in]{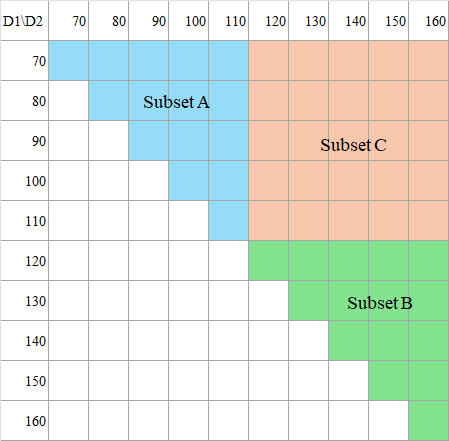}
	\caption{Visual representation of the three subsets. The samples in the low triangle (white) of the matrix are redundant due to the symmetry of the problem. Subsets A (blue) and B (green) are fully simulated and determine the coefficients of the equation. Samples from subset C (orange) have little effect on the determination of the equation coefficients, so they are omitted in FEM simulations. It is shown that the equation can approximate the inductance accurately, even in this dimension range.}
	\label{Fig:SubSets}
\end{figure}

In Fig. \ref{Fig:SubSets} the three subsets A, B, and C are visualized in blue, green and orange, respectively, to clarify their borders. It should be noted that all datasets share the same values for the parameters $w, s, N_T, N_L$, and $O$. They only differ at the values of the outer-side lengths $D_1$ and $D_2$, and consequently at the values of the inner sides $d_1$ and $d_2$. Due to the restriction $d_1, d_2 \ge 17 \text{mm}$ Dataset A is smaller than B, containing 1800 and 4050 samples, respectively.
From the experimental results presented in the following section (Table \ref{Table:EqLab2}) it is demonstrated that the proposed equation can accurately estimate the inductance of windings with dimensions outside the simulation training set as well.

The new equations takes the form

{\setlength{\mathindent}{0cm}
\begin{equation}
	\label{Eq:mn_new}
	L_{\text{MN}} = \alpha_0 \mu_0 D_1^{\alpha_1} D_2^{\alpha_2} \overline{D}_1^{\alpha_3} \overline{D}_2^{\alpha_4} w^{\alpha_5} s^{\alpha_6} N_T^{\alpha_7} N_L^{\alpha_8} O^{\alpha_9(N_L-1)}
\end{equation}

\noindent
where $\overline{D_1} = \frac{D_1+d_1}{2}$, $\overline{D_2} = \frac{D_2+d_2}{2}$. The term $(N_L-1)$ is used to overcome the need for a separate equation in the case of single-layer windings, where $O$ is undefined. As it will be discussed later, the restriction $D_1\leq D_2$, $\overline{D_1} \leq \overline{D_2}$ should be considered.

Applying $\log_{10}$ in both sides of (\ref{Eq:mn_new}) results in the following linear equation

\begin{align}
	\label{Eq:mn_log}
	\log_{10}(L_{\text{MN}}) = c_0 & + \alpha_1 \log_{10}(D_1) + \alpha_2 \log_{10}(D_2) \nonumber \\ 
	&+ \alpha_3 \log_{10}(\overline{D}_1) + \alpha_4 \log_{10}(\overline{D}_2) \nonumber \\
	&+ \alpha_5 \log_{10}(w) + \alpha_6 \log_{10}(s) \nonumber \\
	&+ \alpha_7 \log_{10}(N_T) + \alpha_8 \log_{10}(N_L) \nonumber \\
	&+ \alpha_9(N_L-1) \log_{10}(O)
\end{align} 

\noindent
where $c_0 = \log_{10}(\alpha_0) + \log_{10}(\mu_0)$. Eq. (\ref{Eq:mn_log}) is a linear equation with the form $y = c_0 + c_1 x_1 + c_2 x_2 + \dots + c_9 x_9$. 

Generally, in MLR algorithms ($y = \sum_{i=0}^{N}(c_i x_i)$) the dataset is split into two subsets: one for the calculation of the coefficients $c_i$ (training) and the other to assess the behavior of the equation to samples it has not been exposed to (evaluation). In this case, 80\% of the samples (4680 samples) from the Datasets A and B (as given in Table \ref{Table:Datasets}) comprise the training subset, and the remaining 20\% of the samples (1170 samples) comprise the evaluation subset. Each sample is randomly assigned to either of the subsets, by a pseudo-random algorithm which is a common practice in MLR applications. This process is repeated several times to ensure that the values of the coefficients are converging. The actual results present a small deviation in the vicinity of the nominal values that are presented in Table \ref{Table:EqCoefs}. 

\begin{table}[!tb]
	\centering
	\caption{Monomial Equation Coefficients}
	\label{Table:EqCoefs}
	\begin{tabular}{lcrl}
		\toprule
		Variable 			& Coefficient 	 & \multicolumn{1}{c}{Value}  &  \\ \midrule
		constant 			& $a_0$          & 1.602  &  \\ 
		$D_1$       		& $a_1$          & -0.592 &  \\
		$D_2$       		& $a_2$          & -0.378 &  \\
		$\overline{D}_1$    & $a_3$          & 1.175  &  \\
		$\overline{D}_2$    & $a_4$          & 1.072  &  \\
		$w$        			& $a_5$          & -0.183 &  \\
		$s$        			& $a_6$          & -0.011 &  \\
		$N_T$       		& $a_7$          & 1.794  &  \\
		$N_L$       		& $a_8$          & 1.804  &  \\
		$O^{(N_L-1)}$   	& $a_9$          & -0.006 &  \\ \bottomrule 
	\end{tabular}
\end{table}

The resulting coefficients of the training process $\alpha_i$ are presented in Table \ref{Table:EqCoefs}, along with their corresponding variable (geometrical parameter). The asymmetry that is caused by the restriction $D_1 \le D_2$ is pronounced by the fact that the exponents of $D_1, D_2$ and $\overline{D_1}, \overline{D_2}$ are not equal, and more specifically $|\alpha_1| > |\alpha_2|$ and $\alpha_3 > \alpha_4$. So, for the equation to be valid the restriction $D_1 \le D_2$ should be followed. 

Several metrics can be used to determine the accuracy of the new equation. In Fig. \ref{Fig:error_hist} the histogram of the error 

{\setlength{\mathindent}{2cm}
\begin{equation*}
	\label{Eq:Error}
	\text{error}\% = \frac{L_{SIM}-L_{MN}}{L_{SIM}} \cdot 100
\end{equation*}

\noindent
is presented, where $L_{SIM}$ and $L_{MN}$ are the simulated and estimated inductance, respectively. The histogram is close to a normal distribution with mean value \textmu{} = 0\% and standard deviation \textsigma{} = 1.77\%. The total number of samples with error greater that 5\% is presented in Fig. \ref{Fig:error_group}, grouped according to the number of their layers. This is done in order to ensure that the equation is not biased against any subset of the original dataset. Another reliable metric is the resulting Mean Absolute Error (MAE), as given by

{\setlength{\mathindent}{0.5cm}
\begin{equation}
	\label{Eq:MAE}
	\text{MAE} = \frac{1}{S} \sum_{i=0}^{S} \left( \frac{|L_{SIM}-L_{MN}|}{L_{SIM}} \right) \cdot 100 = 1.24\%,
\end{equation}

\noindent
where $S$ is the number of evaluation samples.

\begin{figure}[!h] 
	\centering
	\subfloat[\label{Fig:error_hist}]{%
		\includegraphics[width=3.5in]{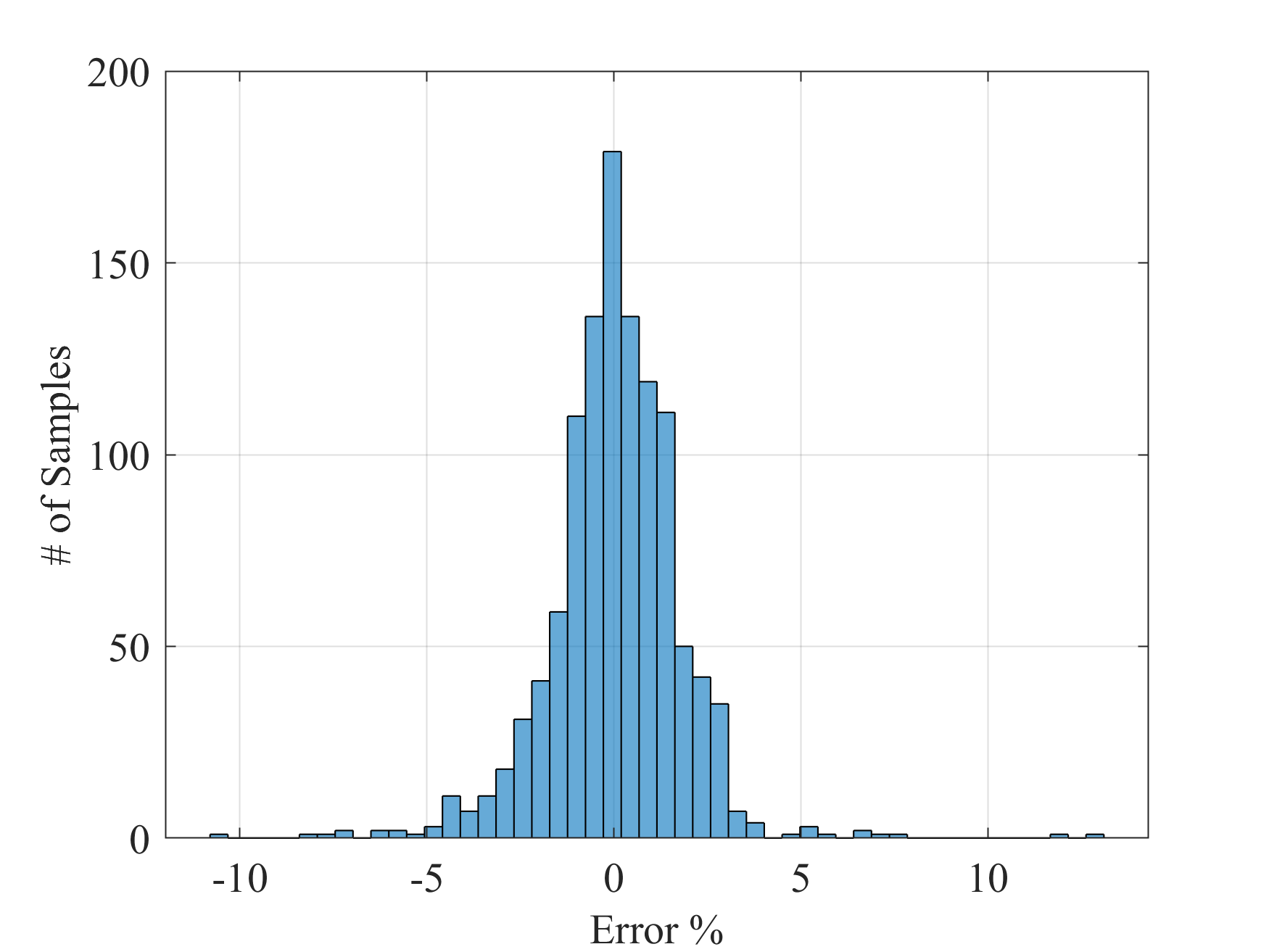}}
	\\
	\subfloat[\label{Fig:error_group}]{%
		\includegraphics[width=3.5in]{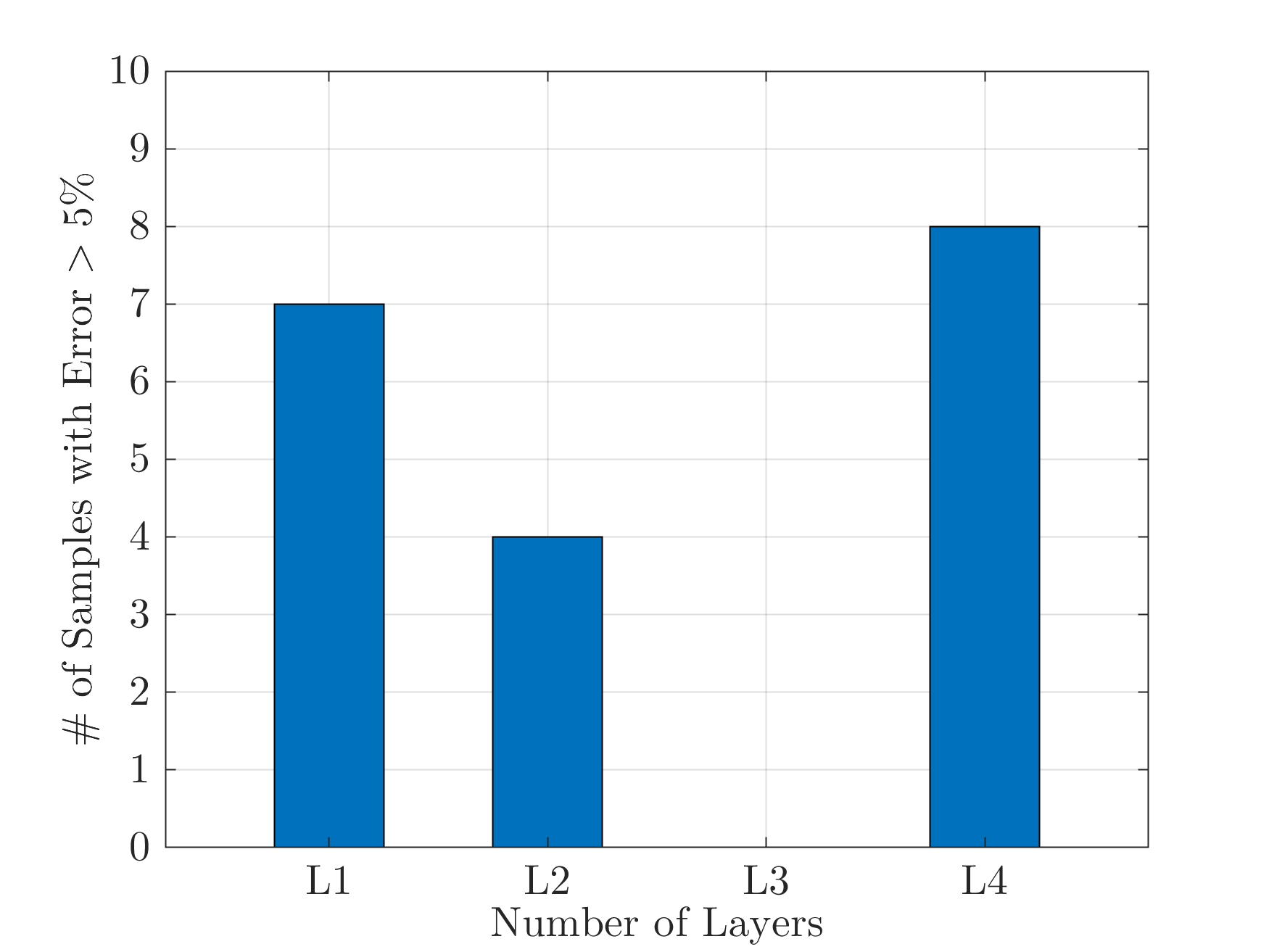}}
	\caption{(a) The total number of samples grouped by error \% and (b) the total number of samples with error > 5\% grouped by the number of layers.}
	\label{Fig:error} 
\end{figure}

In the case of square-shaped windings, where $D_1 = D_2 = D$ and $d_1 = d_2 = d$, \eqref{Eq:mn_new} simplifies to

{\setlength{\mathindent}{0.5cm}
\begin{equation}
	\label{Eq:mn_new_sq}
	L_{\text{MN}} = \alpha_0 \mu_0 D^{\beta_1} \overline{D}^{\beta_2} w^{\alpha_5} s^{\alpha_6} N_T^{\alpha_7} N_L^{\alpha_8} O^{\alpha_9(N_L-1)},
\end{equation}

\noindent
where $\beta_1 = \alpha_1 + \alpha_2 = -0.97$ and $\beta_2 = \alpha_3 + \alpha_4 = 2.247$. In this case, the coefficients of (\ref{Eq:mn_new_sq}) are reasonably close to the corresponding ones in (the original) equation as described in \eqref{Eq:mn}.

Regarding the number of turns per layer $N_T$ and the number of layers $N_L$, their coefficients are practically identical and equal to 1.8. Also, the spacing $s$ seem to have little effect on the overall inductance (less than 2\%), due to the combination of $s$ and $\alpha_6$ values. Hence, (\ref{Eq:mn_new}) can be simplified to 

{\setlength{\mathindent}{0cm}
\begin{align}
	\label{Eq:mn_new_simple}
	L_{\text{MN}} = &1.7274 \mu_0 D_1^{-0.592} D_2^{-0.378} \overline{D}_1^{1.175} \overline{D}_2^{1.072} 
	\nonumber \\
	&w^{-0.183} (N_T N_L)^{1.8} O^{-0.006(N_L-1)},
\end{align}

\noindent
which, is used provides a mean error \textmu{} = 0\%, a standard deviation \textsigma{} = 1.97\%, and an MAE = 1.43\%, slightly higher compared to \eqref{Eq:mn_new}. 

To highlight the usefulness of the approach to estimate the inductance of a winding using a simple (monomial-like) equation, an application is presented in the Appendix. Given a set of dimensional (geometrical) restrictions, an optimization algorithm can be used to maximize the inductance.


\section{Experimental Verification} \label{Sec:ExpVerif}

To confirm the validity of the derived equation in practical windings, ten samples of different geometrical dimensions are examined. The experimental setup, as it is presented in Fig. \ref{fig:JLCLab}, consists of a modified power amplifier, capable of producing 30 V peak, for current and frequency up to 1.5 A peak and 50 kHz, respectively. The parasitic capacitance is small enough to not affect the impedance in any measurable way for frequencies up to 200 kHz.
The impedance of the winding is calculated by correlating the excitation voltage to the amplitude and phase of the current response, and verified with an HP-4284A high-precision LCR meter.

\begin{table*}[h]
	\centering
	\begin{threeparttable}
	\caption{Simulation and Experimental Results}
	\label{Table:EqLab2}
	\begin{tabular}{llrrrrrrrrrrrr}
		\toprule
		& \multicolumn{7}{c}{Dimensions {[}mm{]}}  &    &    & \multicolumn{2}{c}{Inductance {[}\textmu H{]}} &          \\ \midrule
		& & $D_1$  	& $D_2$ 	& $d_1$ 	& $d_2$    	& $w$ 	& $s$   & $O$   & $N_T$ & $N_L$ & $L_{MN}$ & $L_{LAB}$  & Error \% \\
		\multicolumn{1}{c}{\multirow{4}{*}{\rotatebox[origin=c]{90}{Square}} \rdelim\{{4}{0.5mm} } & \#1 & 100.0 	& 100.0 	& 44.0  	& 44.0    	& 4.0 	& 2.0   & 1.60 	& 5  	& 2  	& 9.69     & 9.38    	& 3.31     \\
		\multicolumn{1}{c}{} & \#2 & 100.0 	& 100.0 	& 42.0  	& 42.0    	& 5.0 	& 1.0   & 1.60 	& 5  	& 4  	& 34.09    & 33.51      & 1.71     \\
		\multicolumn{1}{c}{} & \#3 & 100.0 	& 100.0 	& 42.0  	& 42.0    	& 5.0 	& 1.0   & 1.60 	& 5  	& 2  	& 9.09     & 8.91       & 1.95     \\
		\multicolumn{1}{c}{} & \#4 & 100.0 	& 100.0 	& 42.0  	& 42.0    	& 5.0	& 1.0   & 3.20 	& 5  	& 2  	& 9.05     & 8.53       & 6.08     \\
		\multicolumn{1}{c}{\multirow{6}{*}{\rotatebox[origin=c]{90}{Rectangle}} \rdelim\{{6}{0.5mm} } & \#5 & 100.0 	& 163.0 	& 31.0  	& 94.0    	& 3.0 	& 0.5	& -   	& 10 	& 1  	& 13.74    & 13.48      & 1.97     \\
		\multicolumn{1}{c}{} & \#6 & 210.0 	& 294.0 	& 101.0 	& 185.0   	& 5.0 	& 0.5 	& 1.50 	& 10 	& 2  	& 125.70   & 120.80     & 4.05     \\
		\multicolumn{1}{c}{} & \#7 & 120.0 	& 160.0 	& 33.0  	& 73.0    	& 5.0 	& 0.5 	& -   	& 8  	& 1  	& 8.19     & 8.26       & -0.77    \\
		\multicolumn{1}{c}{} & \#8 & 120.0 	& 160.0 	& 33.0  	& 73.0    	& 5.0 	& 0.5 	& 1.60 	& 8  	& 2  	& 29.65    & 30.87      & -3.95    \\
		\multicolumn{1}{c}{} & \#9 & 120.0 	& 160.0 	& 33.0  	& 73.0    	& 5.0 	& 0.5 	& 1.60 	& 8  	& 3  	& 63.87    & 66.40      & -3.81    \\
		\multicolumn{1}{c}{} & \#10 & 100.0 	& 165.0 	& 38.2 		& 103.2 	& 3.0 	& 0.1 	& 0.45 	& 10 	& 4  	& 215.55   & 222.57     & -3.15    \\ \bottomrule 
	\end{tabular}
	
	\smallskip
	\scriptsize
	\begin{tablenotes}
		\RaggedRight
		\item Samples \#7 -\#9 belong to the initial dataset. The other samples have at least one of their parameters outside the dataset. Sample \#10 is the commercial winding. 
	\end{tablenotes}
	
	\end{threeparttable}
\end{table*}

Nine of the windings were printed and folded properly in the lab. Custom printing is preferred to ensure that in multilayer design the distance $O$ between each layer is the same. In commercially printed multilayer boards, copper layers are typically placed near the surface of the board, leaving larger vertical separation in the middle. The exact widths of copper layers, prepreg and core depend on the PCB manufacturer. 
In this case a 10th sample was ordered by a commercial manufacturer. Its values are measured and presented in Fig. \ref{fig:JLCCross}, and are in good agreement with the specifications given by the manufacturer.

The resulting inductances and the error with respect to \eqref{Eq:mn_new} are presented in Table \ref{Table:EqLab2}. It must be noted that even though most of the experimental windings do not belong to the training and evaluation datasets (at least one geometrical parameter is outside datasets A and B), there is only one where the equation estimates an inductance value with 6\% error, compared to the experimentally measured one and all the others present significantly lower errors. Especially for the case of the commercial MLRPW, if $O$ is replaced by the mean value of the prepreg and core, i.e., $(2 \cdot 0.17 + 1)/3 = 0.45$ mm, the equation provides very accurate estimation with only 3.15\% error from the measured value.

\begin{figure}[htb] 
	\centering
	\subfloat[\label{fig:JLCLab}]{%
		\includegraphics[width=3.5in]{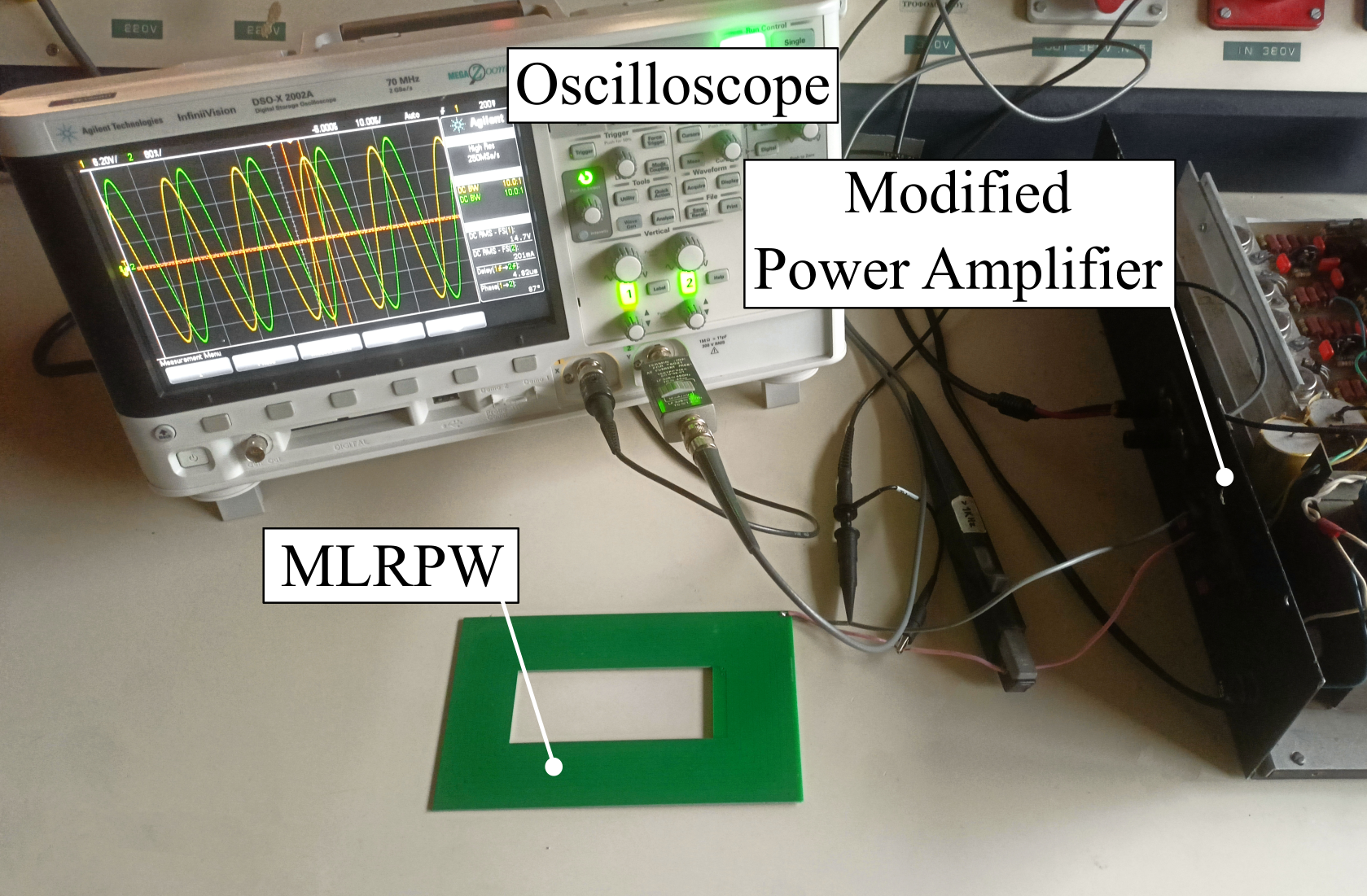}}
	\\
	\subfloat[\label{fig:JLCCross}]{%
		\includegraphics[width=3.5in]{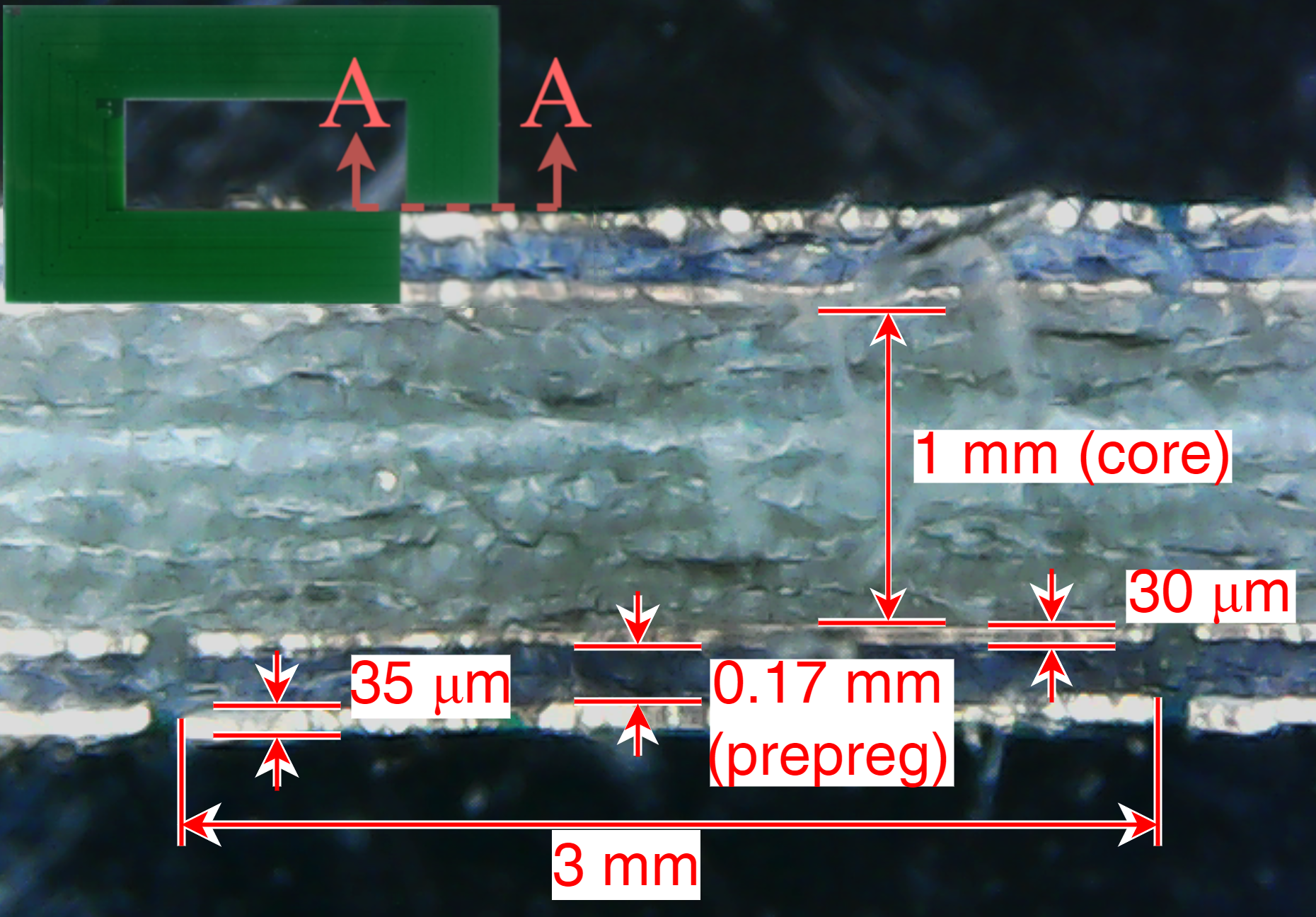}}
	\caption{(a) Laboratory setup for inductance measurement of the MLRPWs at frequency of 50 kHz and current up to 1 A and (b) Cross-section AA of 4-Layer commercial PCB $D_1100D_2165w3s0.1N_T10N_L4O0.45$. Copper layers are not spaced in equal distances along the PCB, but are close to the surface with large spacing in the center.}
	\label{fig:JLC} 
\end{figure}


\section{Conclusions} \label{Sec:Conclusion}

In this study a new, data-fitted via multiple linear regression, monomial-like equation is introduced, which can accurately estimate the inductance of Multilayer Rectangle-shaped Planar Windings for high-power applications. The parameters (geometrical dimensions) are discussed, regarding their role and weight on the inductance. Two datasets, containing approximately 6,000 samples with variable outer-side lengths, trace width, horizontal and vertical spacing, number of turns and layers, are split into two 80/20 training/evaluation subsets, to enable the regression algorithm and the evaluation process. The mean error value and standard deviation are \textmu{} = 0.0\% and \textsigma{} = 1.77\%, respectively, and the MAE = 1.24\%. The equation is not biased against any $N_L$ subgroup of the initial datasets. Finally, the estimations of the equation are compared to laboratory measurements for ten practical windings, verifying the accuracy of the estimations even for windings with dimensions outside the defined dataset.

\bibliographystyle{ieeetr}
\bibliography{refs.bib}

\appendix[Non-Linear Optimization: A Case Study]\label{Sec:Opti}

To highlight the usefulness of closed-form equations, like the one presented in \eqref{Eq:mn_new}, in the design of the winding, an optimization process is presented. The goal is to maximize the inductance considering specific restrictions for the outer- and inner-side lengths, as well as the width and the spacing. The lower and upper boundaries for the different dimensions are presented in Table \ref{Table:optimization}.

\begin{table}[!tbh]
	\centering
	\caption{Lower and upper boundaries for the optimization process.}
	\label{Table:optimization}
	\begin{tabular}{lrrr}
		\toprule
		& \multicolumn{1}{c}{\begin{tabular}[c]{@{}c@{}}lower\\ boundary\end{tabular}} & \multicolumn{1}{c}{\begin{tabular}[c]{@{}c@{}}upper\\ boundary\end{tabular}} & \multicolumn{1}{c}{units} \\ 	\midrule
		$D_1$ & 12.0       & 54.0    & mm  \\
		$D_2$ & 55.0       & 101     & mm  \\
		$d_1$ & 10.5       & 52.0    & mm  \\
		$d_2$ & 54.0       & 99.0    & mm  \\
		$w$  & 2.5        & 5.0     & mm  \\
		$s$  & 0.1        & 1.0     & mm	\\
		\bottomrule
	\end{tabular}
\end{table}

\noindent
The optimization equation is formulated as

\begin{align}
	\max (L_{MN}) &= \min \left(-L_{MN}\right) = \nonumber \\
	\min ( - &1.602 \mu_0 D_1^{-0.592} D_2^{-0.378} \overline{D}_1^{1.175} \overline{D}_2^{1.072}  \nonumber \\
	&w^{-0.183} s^{-0.011} N_T^{1.794} N_L^{1.804} O^{-0.006(N_L-1)} ),
\end{align}

\noindent
subject to:

{\setlength{\mathindent}{4cm}
	\begin{equation} \label{Eq:Linear_Constrains}
		D_1 < D_2,	
\end{equation}}


{\setlength{\mathindent}{2cm}
	\begin{equation} \label{Eq:nonLinear_Constrains1}
		d_1 = D_1 - 2 N_T (w + s) + 2s,
\end{equation}}

{\setlength{\mathindent}{2cm}
	\begin{equation} \label{Eq:nonLinear_Constrains2}
		d_2 = D_2 - 2 N_T (w + s) + 2s,
\end{equation}}

{\setlength{\mathindent}{2cm}
	\begin{equation} \label{Eq:nonLinear_Constrains_N}
		N_T \in \{3, 4, \dots 9, 10\} \subset \mathbb{N}^*.
\end{equation}}
s
\noindent
Each variable should receive values inside the interval that is defined from the lower and upper boundaries, as they are presented in Table \ref{Table:optimization}.

The linear restriction \eqref{Eq:Linear_Constrains} arises from the asymmetry of the training dataset, as it is explained in Section \ref{Sec:EqForm}. Furthermore, the non-linear restrictions \eqref{Eq:nonLinear_Constrains1} and \eqref{Eq:nonLinear_Constrains2} result from the fact that the variables of the equation represent the geometrical parameters of the windings, which depend on each other. Finally, the non-linear restriction \eqref{Eq:nonLinear_Constrains_N} arises from the fact that the number of turns shall be an integer number.

The number of layers $N_L$ is not considered in this optimization process, as it is independent of the other variables and as it increases, so does the total inductance. Similarly, the vertical distance between two consecutive layers $O$ is not considered, since it depends on the production capabilities of the PCB manufacturer, and its increase leads to inductance reduction.

The problem is characterized as non-linear mixed integer optimization, which can be solved by predefined optimization functions of Matlab. In this case the \texttt{fmincon} is selected.

Furthermore, the problem is non-convex in the general case, and may lead to several local minima. One way to overcome this issue is running the optimization algorithm for different initial points, which can be randomized for each iteration. 

For each iteration the optimization algorithm converges to a specific state vector $\vec{x_{i}} = [D_1 ~ D_2 ~ d_1 ~ d_2 ~ w ~ s ~ N_T]$ and returns a local maximum of $L_{MN,i}$. When the $L_{MN,i}$ is greater than the temporary $L_{MN,opt}$, and the constraints are not violated, the temporary $\vec{x_{opt}}$ is updated to the current $\vec{x_{opt}} = \vec{x_{iter}}$ and $L_{MN,opt} = L_{MN,i}$. 

It must be noted that this method does not mathematically guarantee that the global maximum is achieved. However, practically, running the algorithm for a large number of iterations, provides a very good solution. The optimal results for the conditions of Table \ref{Table:optimization} are presented in Table \ref{Table:optimization_results}. 

\begin{table}[!tbh]
	\centering
	\caption{Optimization algorithm results.}
	\label{Table:optimization_results}
	\begin{tabular}{cccccccc}
		\toprule
		\multicolumn{6}{c}{{[}mm{]}}          & turns & {[}\textmu H{]} \\ 	
		$D_1$   & $D_2$  & $d_1$   & $d_2$   & $w$    & $s$   & $N_T$    & $L$        \\ \midrule
		\multicolumn{1}{r}{54.0} & \multicolumn{1}{r}{101} & \multicolumn{1}{r}{12.6} & \multicolumn{1}{r}{59.6} & \multicolumn{1}{r}{2.5} & \multicolumn{1}{r}{0.1} & \multicolumn{1}{r}{8} & \multicolumn{1}{r}{63.82}	\\	\bottomrule
	\end{tabular}
\end{table}

It can be noted that the $\max (L)$ is achieved for $\max(D_1)$ and $\max(D_2)$, but not for $\max(d_1)$, $\max(d_2)$, or $\max(N_T)$.
A winding of these dimensions has been designed and commercially printed, with a measured inductance of \mbox{63.43 \textmu H}, resulting in 0.7\% error compared to the equation.

\end{document}